\documentclass[12pt]{article}
\usepackage{epsf}

\begin{document}


\newcommand\balpha{\mbox{\boldmath $\alpha$}}
\newcommand\bbeta{\mbox{\boldmath $\beta$}}
\newcommand\bgamma{\mbox{\boldmath $\gamma$}}
\newcommand\bomega{\mbox{\boldmath $\omega$}}
\newcommand\blambda{\mbox{\boldmath $\lambda$}}
\newcommand\bmu{\mbox{\boldmath $\mu$}}
\newcommand\bphi{\mbox{\boldmath $\phi$}}
\newcommand\bzeta{\mbox{\boldmath $\zeta$}}
\newcommand\bsigma{\mbox{\boldmath $\sigma$}}
\newcommand\bepsilon{\mbox{\boldmath $\epsilon$}}

\newcommand{\be}{\begin{eqnarray}}
\newcommand{\ee}{\end{eqnarray}}
\newcommand{\nn}{\nonumber}

\newcommand{\ft}[2]{{\textstyle\frac{#1}{#2}}}
\newcommand{\eqn}[1]{(\ref{#1})}
\newcommand{\vsone}{\vspace{1cm}}
\newcommand{\para}{\paragraph{}}
 
\begin{titlepage}

\begin{flushright}
hep-th/0005186\\
MIT-CTP-2985\\
KCL-TH-00-28\\
\end{flushright}
\begin{centering}
\vspace{.2in}
{\Large {\bf Dynamics of N=2 Supersymmetric Chern-Simons Theories}}\\
\vspace{.4in}
David Tong${}$ \\
\vspace{.4in}
Center for Theoretical Physics\\
Massachusetts Institute of Technology,\\
Cambridge MA 02139, USA.\\
\vspace{0.1in}
and\\
\vspace{0.1in}
Department of Mathematics\\
Kings College, The Strand, \\ 
London, WC2R 2LS, UK\\
\vspace{.05in}
{tong@mth.kcl.ac.uk}\\
\vspace{.6in}
{\bf Abstract} \\
\end{centering}
\vspace{0.2in}
We discuss several aspects  of three dimensional ${\cal N}=2$ 
supersymmetric gauge theories coupled to chiral multiplets. 
The generation of Chern-Simons couplings at low-energies 
results in novel behaviour including compact Coulomb branches, 
non-abelian gauge symmetry enhancement and interesting patterns 
of dynamically generated potentials. 
We further show how, given any pair of mirror theories 
with ${\cal N}=4$ supersymmetry, one may flow to 
a pair of mirror theories with ${\cal N}=2$ supersymmetry 
by gauging  a suitable combination of the R-symmetries. 
The resulting theories again have interesting properties due 
to Chern-Simons couplings. 

\vspace{.1in}


\end{titlepage}
\section{Introduction}

Many three dimensional gauge theories with ${\cal N}=4$ supersymmetry 
exhibit mirror symmetry, a phenomenon in which two theories with different 
ultra-violet descriptions flow to the same infra-red physics. 
Initially discovered by Intriligator and Seiberg \cite{intseib}, 
many further pairs of mirror theories have since been constructed 
using various methods \cite{berk1,berk2}. Moreover, Kapustin and Strassler have   
suggested an elegant 
formula which captures many aspects of mirror symmetry in ${\cal N}=4$ 
abelian gauge theories \cite{ks}. 

Under the mirror map, the two $SU(2)$ R-symmetry groups of the three 
dimensional theories are exchanged, together with the mass and Fayet-Iliopoulos 
(FI) parameters, and the Coulomb and Higgs branches. This latter exchange 
is particularly interesting as the Higgs branch metric is 
protected against quantum corrections, while the Coulomb branch 
metric may receive not only one-loop perturbative corrections, but also 
corrections from instantons, loops around the background of instantons 
and, more surprisingly, instanton-anti-instanton pairs. In the 
mirror theory, all of these effects are captured by a classical, hyperK\"ahler 
quotient construction of the Higgs branch.

As with all dualities, it is natural to ask if one can continue to 
make progress with less supersymmetry. Indeed, several pairs of 
mirror theories with ${\cal N}=2$ and, more speculatively, ${\cal N}=1$ 
supersymmetry have been 
proposed \cite{berk3, ahiss, dt, katz}. In particular, 
the authors of \cite{ahiss} give a prescription for constructing such models given an 
${\cal N}=4$ pair. One first adds an ${\cal N}=2$ multiplet to one theory, explicitly 
breaking half of the supersymmetry. If the scalar in this multiplet can be 
interpreted as dynamical mass parameter for some of the fields, then one 
may determine the mirror deformation: it must play the role of a dynamical 
FI parameter.

The first part of this paper will be concerned with providing an 
alternative 
prescription for flowing to ${\cal N}=2$ theories. The basic 
observation is very simple: there exists a combination of R-symmetries 
that may be gauged without fully breaking supersymmetry. Weakly gauging 
this symmetry introduces an axial mass, allowing a subset of the chiral 
multiplets to be integrated out. For abelian ${\cal N}=4$ theories, one 
finds that the resulting ${\cal N}=2$ theories have Chern-Simons couplings 
and several novel features, including the possibility of compact Coulomb 
branches. The abelian mirror pairs have been 
previously given in \cite{dt} where it was shown, using techniques of 
toric geometry, that the Higgs and Coulomb branches coincide. The method 
introduced here reproduces the results of \cite{dt}, and may also 
be applied to theories with non-abelian gauge symmetry. 

In section 3, we discuss the low-energy dynamics of ${\cal N}=2$ non-abelian 
theories with chiral 
multiplets. Perturbatively, the Coulomb branches exhibit similar 
behaviour to the abelian theories and, in particular, are compact. 
However instanton effects generate a superpotential which drives the vacuum 
to the boundary of 
the perturbative Coulomb branch.  For $SU(N)$ gauge groups with $N\geq 3$, 
this results in  dynamical supersymmetry breaking.  
Section 4 summarises the main points.

\section{Gauging the R-symmetry: Abelian Theories}

We start by considering abelian theories. Later in this 
section, we will describe the most general abelian ${\cal N}=4$ 
mirror pairs, together with the deformation that breaks 
supersymmetry to ${\cal N}=2$. However, we first illustrate these 
ideas with a simple example which will also serve to set our notations and 
conventions.  

\subsection*{The Self-Mirror Theory}

The simplest example of mirror symmetry in three dimensions 
is the ${\cal N}=4$ self-mirror theory of Intriligator and Seiberg 
\cite{intseib}. 
It consists of a $U(1)_G$ vector multiplet, together with two hypermultiplets. 
As we intend to partially break supersymmetry in the very near future, 
let us resort to four supercharge notation from the off. The 
${\cal N}=4$ vector multiplet consists of an ${\cal N}=2$ vector 
multiplet $V$ together with an ${\cal N}=2$ chiral multiplet $\Psi$. 
We denote the real scalar in the former as $\phi$ and the complex 
scalar in the latter as $\psi$. Together, these scalars transform 
in the $({\bf 3},{\bf 1})$ of the $SU(2)_N\times SU(2)_R$ R-symmetry. 

Each of the $i=1,2$  hypermultiplets consists of two chiral multiplets, $Q_i$ and 
$\tilde{Q}_i$. These contain the 
complex scalars  
$q_i$ and $\tilde{q}_i$ which have charge $+1$ and $-1$ respectively. 
These scalars transform in the $({\bf 1},{\bf 2})$ representation 
of the R-symmetry.

The theory has two further global symmetries. Firstly there exists an 
$SU(2)_F$ flavour symmetry which acts in the obvious fashion on the 
hypermultiplet index $i=1,2$. Weakly gauging this symmetry introduces a 3-vector 
of mass parameters, ${\vec m}$. These transform in the same manner as the 
vector multiplet scalars under the R-symmetry and, without loss 
of generality, we may use this freedom to set ${\vec m}=(0,0,m)$. This 
component of the mass vector is commonly referred to as a real mass; it 
is the complex mass that has been set to zero.

The second global symmetry is less obvious in the Lagrangian formalism as 
it requires a dualisation to magnetic variables. Recall that in three 
dimensions one may exchange the photon in preference of a periodic scalar $\sigma$.  
Shifting this scalar by a constant, $\sigma\rightarrow\sigma + c$, 
is a symmetry of the theory 
at the classical and perturbative level. For abelian theories there 
are no instanton effects and this symmetry, which is usually denoted 
as $U(1)_J$, survives in the full quantum theory. Once again, one may consider 
weakly gauging this symmetry\footnote{This current couples to a twisted vector 
multiplet through a BF-coupling \cite{ks}} and the result is the introduction of 
a 3-vector of FI parameters ${\vec \zeta}$, transforming as 
$({\bf 1},{\bf 3})$ under the R-symmetry. The R-symmetry may again be 
employed to  rotate the FI parameters to be 
${\vec \zeta}=(0,0,\zeta)$. 

Notice in particular that in 
the presence of both mass and FI parameters, the R-symmetry is 
broken to its Cartan subalgebra, $U(1)_N\times U(1)_R$, while the 
flavour symmetry is similarly broken to $U(1)_F$. It will 
prove useful in what follows to document the transformation 
of the hypermultiplet scalars under these various symmetries
\be
\begin{array}{cccccc} {} &{} & q_1 & \tilde{q}_1 & q_2 & \tilde{q}_2 \\
\\
U(1)_G & & +1 & -1 & +1 & -1   \\
U(1)_F & & +1 & -1 & -1 & +1   \\
U(1)_R & & +1 & +1 & +1 & +1 
\end{array}
\label{youwons}\ee
The scalar potential with both masses and FI parameters non-zero is 
given by,
\be
U&=&e^2(|q_i|^2-|\tilde{q}_i|^2-\zeta)^2 
+e^2|\tilde{q}_iq_i|^2 +|\psi|^2(|q_i|^2+|\tilde{q}_i|^2) \nn\\ 
&&\ +(\phi+m)^2|q_1|^2+(-\phi-m)^2|\tilde{q}_1|^2
\label{pot1}\\ 
&&\ +(\phi-m)^2|q_2|^2 
+(-\phi+m)^2|\tilde{q}_2|^2
\nn\ee
where $e$ is the gauge coupling constant which has dimension of (mass)${}^{1/2}$ 
and summation over $i$ is 
assumed in the first three terms. The final four terms are the real masses 
for each of the chiral multiplets. The $\pm$ signs in front of $\phi$ and $m$ 
reflect charges \eqn{youwons} 
of the scalars under $U(1)_G$ and $U(1)_F$ respectively. 
Importantly, ${\cal N}=4$ supersymmetry ensures that  
$Mass(q_i)=-Mass(\tilde{q}_i)$. 

The statement of mirror symmetry in this theory is that the Higgs branch 
and Coulomb branch metrics coincide \cite{intseib}. Let us review how this 
comes about. For vanishing masses, the space of vacua is given by $\phi=\psi=0$ 
while the hypermultiplet scalars are restricted by the first two terms 
in \eqn{pot1} modulo $U(1)_G$. The result is that the 
theory possesses a 4-dimensional Higgs branch with the Eguchi-Hanson metric 
given naturally by a hyperK\"ahler quotient construction. The 
two-sphere which sits in the middle of this space has size $\zeta$. 
This metric is classically exact: it receives no quantum corrections \cite{aps}.

In contrast, when $\zeta=0$, the space of vacua is given by 
$q_i=\tilde{q}_i=0$, while $\phi$, $\psi$ and the dual photon $\sigma$ are 
unconstrained, resulting in a 4-dimensional Coulomb branch. 
While classically flat, the metric on this branch does receive quantum 
corrections. After integrating out the hypermultiplets at one 
loop, one finds the resulting metric is double-centered Taub-NUT 
space which, in the limit $e^2\rightarrow\infty$ becomes once again 
Eguchi-Hanson. The two-sphere in the center has size $m$. 
There are no higher loop corrections to this metric. 

Thus this theory is self-mirror \cite{intseib}: in the infra-red limit 
$e^2\rightarrow\infty$, the low-energy physics on the Coulomb branch 
is the same as that on the Higgs branch {\it if} one 
swaps mass and FI parameters and $SU(2)_N$ and $SU(2)_R$ symmetries. 

It is instructive to compare the isometries of the two spaces. These 
descend from the symmetries of the theories. The Higgs branch has an 
$SU(2)_F\times U(1)_R$ symmetry, reflecting the transformation of 
the hypermultiplet scalars discussed earlier. Meanwhile, for 
finite $e^2$, the Coulomb branch has a $U(1)_J\times U(1)_N$ 
isometry. However, in the strong coupling  this 
is enhanced to $SU(2)_J\times U(1)_N$. Such enhancement of global 
flavour symmetries in the infra-red is commonplace on the Coulomb 
branch of mirror pairs and is still rather poorly understood. 

\subsubsection*{\it Gauging the R-Symmetry}

Having reviewed mirror symmetry in this self-mirror case, let us 
proceed to break supersymmetry to ${\cal N}=2$. The basic idea is 
simple: just as we weakly gauged the global flavour and shift 
symmetries above in order to introduce background parameters, we may 
perform a similar operation with the R-symmetry. Of course, 
usually gauging the R-symmetry results in supersymmetry being 
completely broken. 
In order to see this, we need only observe that the 
superpotential of a theory transforms under the R-symmetry and 
therefore it is not possible to write the Lagrangian in a manifestly 
gauge invariant and supersymmetric fashion. In the present case, the 
superpotential is given by
\be
{\cal W}=\tilde{Q}_i\Psi Q_i
\nn\ee
For non-zero mass and FI parameters, the residual R-symmetry group is 
$U(1)_N\times U(1)_R$, under which $Q_i$ and $\tilde{Q}_i$ both have 
charge $(0,+1)$ while $\Psi$ has charge $(+2,0)$. We see that the 
superpotential indeed transforms under each of the R-symmetries. 
However, it is clear that if we choose to gauge only the axial 
combination, $U(1)_{R-N}$, then the superpotential is invariant and 
${\cal N}=2$ supersymmetry will remain manifestly  
unbroken. Moreover, as mirror symmetry exchanges the two R-symmetries, 
the mirror deformation must require the same current to be gauged, 
albeit with the introduction of a minus sign:  $U(1)_{N-R}$.
For the remainder of this section, we will bring 
this simple trick to bear on various theories. For now, let us 
content ourselves with examining the consequences in the 
simple case under discussion. 

In weakly gauging $U(1)_{R-N}$, we introduce a single, 
real\footnote{Previously we gauged the background symmetries in an ${\cal N}=4$ 
supersymmetric fashion resulting in three background parameters,  
reflecting the existence of three real scalars in the ${\cal N}=4$ 
vector multiplet. In the current situation, we wish to preserve ${\cal N}=2$ 
supersymmetry, and thus introduce only a single real background parameter.}
background parameter $X$. From the charges assigned to the various 
fields, one may determine the new scalar potential,
\be
U&=&e^2(|q_i|^2-|\tilde{q}_i|^2)-\zeta)^2 
+e^2|\tilde{q}_iq_i|^2 +|\psi|^2(|q_i|^2+|\tilde{q}_i|^2) + 4X^2|\psi|^2 
\nn\\ &&\ +(\phi+m\pm X)^2|q_1|^2
+(-\phi-m\pm X)^2|\tilde{q}_1|^2 \label{pot2}\\ &&\ +(\phi-m\pm X)^2|q_2|^2 
+(-\phi+m\pm X)^2|\tilde{q}_2|^2
\nn\ee
where the $\pm$ signs in front of the $X$'s depend on whether we 
choose to gauge $U(1)_{R-N}$ (plus sign by convention) or $U(1)_{N-R}$ 
(minus sign).  We will refer to the case with $+X$ as Theory A, and 
the case with $-X$ as Theory B.

The effect of gauging the R-symmetry is two-fold: the chiral 
multiplet $\Psi$ has gained a mass, and a mass splitting has been introduced 
between $Q_i$ and $\tilde{Q}_i$. Although we have exhibited mass terms 
only for scalars, ${\cal N}=2$ supersymmetry ensures that the 
fermionic superpartners have a similar splitting. The claim is that, 
by construction, the theory with deformation $+X$ is mirror to the 
theory with deformation $-X$. Let us now demonstrate this explicitly 
in the limit that $X\rightarrow\infty$. In doing so, we will uncover 
several interesting properties of these theories. 

Firstly, consider Theory A. Clearly we do not want to 
integrate out all fields, so we perform the shift,
\be
\phi^\prime=\phi+X
\nn\ee
and require that $m$, together with the rescaled scalar ${\phi}^\prime$, 
remain finite as $X\rightarrow \infty$. The chiral multiplets $\Psi$, 
$\tilde{Q}_1$ and $\tilde{Q}_2$ thus become heavy and may be 
integrated out. The $\Psi$ field is uncharged under the 
gauge field and simply decouples as its mass becomes infinite. 

However, things are somewhat more interesting for the charged chirals, 
for even in the limit of infinite mass they do not decouple completely. 
In particular, when integrated out, they generate a Chern-Simons 
coupling \cite{redlich,agw},
\be
\frac{\kappa}{4\pi} \epsilon^{\mu\nu\rho}A_\mu F_{\nu\rho} 
\nn\ee
together with its supersymmetric completion. Importantly, as will 
be discussed shortly, this supersymmetric completion includes a 
contribution to the scalar potential \cite{klee}. Consider the more general 
situation in which we have $N$ chiral superfields of charge $s_i$ 
and real mass $M_i$. Integrating out these fields induces a Chern-Simons 
coupling
\be
\kappa = \ft12\sum_{i=1}^N s_i^2\ {\rm sign}(M_i)
\label{csgeneral}\ee
A very nice review of many properties of Chern-Simons theories, including 
a discussion of these dynamically generated terms, can be 
found in \cite{dunne}. Equation \eqn{csgeneral} shows us the importance of 
the previous observation 
that $Mass(q_i)=-Mass(\tilde{q}_i)$. This ensures that in ${\cal N}=4$ 
supersymmetric theories, we may integrate out hypermultiplets 
with impunity without fear of generating CS terms. The same is 
not true of chiral multiplets in ${\cal N}=2$ theories. 

Returning to the example in hand, the dynamically generated 
CS parameter is given by,
\be
\kappa = \ft12 \left({\rm sign} (-\phi^\prime-m+2X)+
{\rm sign} (-\phi^\prime+m+2X)\right) = +1
\nn\ee
At the same time, there is also a finite renormalisation of the 
FI parameter. The simplest way to see this is to notice that 
the FI parameter plays the role of a cross CS term between 
$U(1)_G$ and both $U(1)_F$ and $U(1)_{R-N}$. While the former doesn't 
contribute, the latter gives,  
\be
\zeta \rightarrow \zeta^\prime=\zeta-X
\label{fishift}\ee
And, naturally, we take this rescaled FI parameter to be finite. 
Thus we are left with:

\paragraph{}
{\bf Theory A:} ${\cal N}=2$ $U(1)$ gauge theory with 
two chiral multiplets, both of charge $+1$, with real 
masses $+m$ and $-m$ respectively, a CS parameter $+1$ and FI parameter 
$\zeta^\prime$.

\paragraph{}

It will be instructive to examine the scalar potential 
of this theory,
\be 
U_A= e^2( |q_1|^2 + |q_2|^2 -\kappa\phi - \zeta^\prime)^2
+(\phi+m)^2|q_1|^2+(\phi-m)^2|q_2|^2
\label{pot4}\ee
with $\kappa=+1$. Notice in particular the presence of the term $\kappa\phi$ 
inside the D-term. This is part of the supersymmetric completion 
of the CS term mentioned previously.

We turn now to the mirror theory. Noting that Theory A required 
a shift of the FI parameter \eqn{fishift}, it is natural to 
conjecture that Theory B requires a similar shift of the 
mass parameter, 
\be
m^\prime=m-X
\nn\ee
Now in the limit $X\rightarrow\infty$, with $\phi$ and $m^\prime$ 
kept fixed, we are forced to integrate out $\psi$, $\tilde{q}_1$ and 
$q_2$. Notice that this is a different combination of chiral 
fields from the previous case, which means that while a CS 
parameter is still generated, $\kappa=-1$, the FI parameter 
is not renormalised in this case. Thus we are led to 

\paragraph{}

{\bf Theory B:} ${\cal N}=2$ $U(1)$ gauge theory with 
two chiral multiplets, of charge $+1$ and $-1$, both 
with real masses $+m^\prime$, a CS parameter $-1$ and FI parameter 
${\zeta}$.

\paragraph{}

The scalar potential for this theory is given by
\be
U_B=e^2( |q_1|^2-|\tilde{q}_2|^2-\kappa\phi-\zeta)^2+(\phi+m^\prime)^2
|q_1|^2+(-\phi+m^\prime)^2|\tilde{q}_2|^2
\label{pot5}\ee
with $\kappa=-1$.

It was shown in \cite{dt} that the vacuum moduli spaces of 
Theory A and Theory B do indeed coincide if we exchange 
$m$ with $\zeta$ and $m^\prime$ with $\zeta^\prime$. 
Here we review the basic features of mirror symmetry in these 
theories. Firstly, we consider the Higgs branch of Theory A. 
This exists if $m=0$ and is well known to be a copy of 
${\bf CP}^1$ of K\"ahler class $\zeta^\prime>0$. The challenge 
is to reproduce this as the Coulomb branch of Theory B when 
$\zeta =0$.

The observation of \cite{dt} is that while classically 
the presence of the CS parameter in \eqn{pot5} ensures that 
there is no Coulomb branch, this situation is improved 
by quantum effects. Specifically, let us set $q_1=\tilde{q}_2=0$, 
and $\phi\neq 0$. Then, upon integrating out the remaining 
chiral multiplets, the CS parameter receives a further correction,
\be
\kappa \rightarrow -1 + \ft12\left({\rm sign}(\phi+m^\prime)+
{\rm sign}(-\phi+m^\prime)\right)
\nn\ee
The existence of the Coulomb branch hinges on the vanishing of this 
quantity, which occurs only when $|\phi|\leq m^\prime>0$. 
Thus the dynamical generation 
of CS terms restricts the Coulomb branch to the interval 
$-m^\prime\leq \phi\leq m^\prime$. 

This is the first part of the story. The second part concerns 
the dual photon $\sigma$, which provides us with a circle 
fibered over the interval. It was argued in \cite{ahiss,dt} that  
the end points of the interval must be fixed points of the 
$U(1)_J$ symmetry as there exist extra massless degrees of 
freedom at these points which are invariant under $U(1)_J$. 
The circle therefore shrinks to zero size at the end points 
of the interval. The resulting Coulomb branch is depicted 
in Figure 1. 

\begin{figure}
\begin{center}
\epsfxsize=3.0in\leavevmode\epsfbox{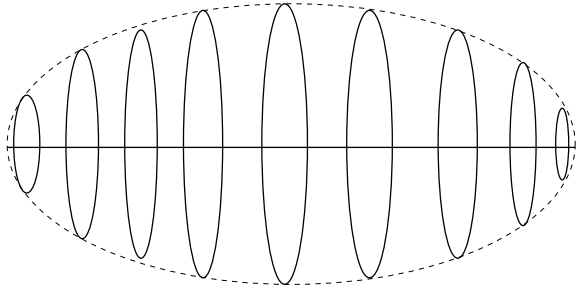}
\end{center}
\caption{\em The Coulomb branch of Theory B. The scalar $\phi$ (plotted 
horizontally) is restricted to lie within the interval $-m<\phi<m$. The dual 
photon provides a further ${\bf S}^1$ which is fibered over this 
interval such that its radius vanishes at the end points, resulting 
in a Coulomb branch with topology of the sphere.}
\label{fig1}
\end{figure}

Classically the metric on the Coulomb branch is given by
\be
{\rm d}s^2 = H{\rm d}\phi^2 + H^{-1}{\rm d}\sigma^2
\nn\ee
where $H=1/e^2$. While the K\"ahler potentials of theories with four 
supercharges are generically not protected against quantum corrections, 
in this theory we may argue that the metric receives only one-loop 
corrections. To see this recall that mirror symmetry requires the  
Coulomb branch to develop an enhanced $SU(2)_J$ isometry group in the 
infra-red which, given the topology, restricts the metric to 
be Fubini-Study. Indeed, at one-loop, we find
\be
H_{one-loop}=\frac{1}{e^2}+\frac{\ft12}{|\phi+m|}
+\frac{\ft12}{|-\phi+m|}=\frac{1}{e^2}+\frac{m}{\phi^2-m^2}
\nn\ee
which becomes Fubini-Study as $e^2\rightarrow\infty$.

\subsection*{General Abelian Theories}

The most general ${\cal N}=4$ abelian mirror pairs were exhibited in 
\cite{berk2}. The following presentation differs somewhat from that 
reference, but may easily be seen to be equivalent. In particular, 
the following formalism of mirror symmetry arises naturally from the 
Kapustin-Strassler formula \cite{ks}. The mirror pairs 
are:

\paragraph{}

{\bf Theory A:} ${\cal N}=4$ $U(1)^r$ gauge theory with $N$ hypermultiplets of 
charge $R_i^a$, $i=1,\cdots,N$ and $a=1,\cdots,r$, where $N\geq r$ 
and the charge matrix is taken to have maximal rank. The theory 
has FI parameters $\vec{\zeta}^a$ and mass parameters $\vec{m}_i$, each 
of which is a 3-vector. Notice 
that only $N-r$ of the mass parameters are independent.

\paragraph{}

{\bf Theory B:} ${\cal N}=4$ $U(1)^{N-r}$ gauge theory with $N$ hypermultiplets 
of charge $S^p_i$, $i=1,\cdots,N$ and $p=1,\cdots,N-r$. Again, the 
charge matrix is taken to be of maximal rank. This theory has vector  
FI parameters $\hat{\vec{\zeta}^p}$ and mass parameters $\hat{\vec{m}}_i$, where 
$N-(N-r)=r$ of the mass parameters are independent.

\paragraph{}

The charges of Theory A and Theory B are constrained to satisfy
\be
\sum_{i=1}^NR_i^aS_i^p=0\ \ \ \ \ \ \ \mbox{for all $a$ and $p$}
\label{charges}\ee
This restriction ensures that the charges $S_i^p$ may be thought 
of as a basis of generators for the Cartan subalgebra of the 
flavour symmetry of Theory A. Likewise, $R^a_i$ provide a basis 
for the abelian flavour symmetry of Theory B. 
The relationship between the mass parameters of Theory A and the 
FI parameters of Theory B is most conveniently described in terms 
of $N$ mirror invariant 3-vectors, $(\vec{n}_a,\hat{\vec{n}}_p)$, through
\be
\vec{\zeta}^a= R^a_iR_i^b\vec{n}_b\ \ \ \ &,&\ \ \ \ 
\vec{m}_i=S_i^p\hat{\vec{n}}_p \nn\\
\hat{\vec{\zeta}^p}= S_i^pS_i^q\hat{\vec{n}}_q\ \ \ \ &,&\ \ \ \ 
\hat{\vec{m}}_i=R_i^a\vec{n}_a
\label{mirrinv}\ee
from which we find the desired relations 
$\vec{\zeta}^a= R_i^a\hat{\vec{m}}_i$ and 
$\hat{\vec{\zeta}^p}= S_i^p\vec{m}_i$. 

The statement of mirror symmetry is that the classical metric on 
the Higgs branch 
of Theory A coincides with the one-loop corrected metric on the Coulomb 
branch of Theory B in the infra-red limit $e^2\rightarrow\infty$ and 
vice-versa, if the masses and FI parameters are related as above. 

For generic values of the mirror-invariants, the $SU(2)_N\times SU(2)_R$ 
R-symmetry of the theory is completely broken. This will not do for 
our purposes. We therefore restrict attention to the subset of parameter 
space in which $\vec{n}_a=(0,0,n_a)$ and $\hat{\vec{n}}_q=(0,0,\hat{n}_q)$, such 
that all complex mass and FI parameters are set to zero. 
This ensures that the residual R-symmetry is once again $U(1)_N\times U(1)_R$. 

Weakly gauging the axial combination $U(1)_{R-N}$ with a background parameter $X$ 
again introduces a mass for each chiral multiplet in the vector multiplets 
of Theory A. As previously, these simply decouple in the limit 
$X\rightarrow\infty$ and the interesting 
physics occurs due to the mass splitting of the hypermultiplets. Specifically, 
the masses of the hypermultiplets of Theory A are given by
\be
\sum_{i=1}^N(R_i^a\phi_a+m_i+X)^2|q_i|^2 
+\sum_{i=1}^N(-R_i^a\phi_a-m_i+X)^2|\tilde{q}_i|^2
\label{pot6}\ee
Once again, we must perform a rescaling of fields before integrating out 
those we deem to be heavy. We choose the rescaling, 
\be
R_i^a\phi_a^\prime&=&R_i^a\phi_a+\ft12 X \nn\\
m^\prime_i&=& m_i+\ft12 X
\label{massnew}\ee
in terms of which, the masses \eqn{pot6} are given by
\be
\sum_{i=1}^N(R_i^a\phi^\prime_a+m^\prime_i)^2|q_i|^2 
+\sum_{i=1}^N(-R_i^a\phi^\prime_a-m_i^\prime+2X)^2|\tilde{q}_i|^2
\label{pot7}\ee
leaving us with the task of decoupling the $\tilde{q}_i$'s.  
The 
induced CS couplings now include the possibility of cross-terms of the form 
$(1/4\pi)\kappa^{ab}\epsilon_{\mu\nu\rho}A^\mu_aF^{\nu\rho}_b$ 
and the generalisation of \eqn{csgeneral} to include such cross terms is given by,
\be
\kappa^{ab}&=&\ft12\sum_{i=1}^NR_i^aR_i^b\ {\rm sign}(M_i)\nn\\
&=& \ft12R_i^aR^b_i
\label{crosskappa}\ee
where $M_i$ is the mass of the $i^{\rm th}$ chiral multiplet which, in 
the present case, is necessarily positive for $\tilde{q}_i$ in the limit 
$X\rightarrow\infty$.  
Similarly, there is a finite renormalisation of the FI parameter,
\be
\zeta^a\rightarrow\zeta^{\prime a}= 
\zeta_a-\ft12\sum_{i=1}^N(X-m_i)R_i^a=\zeta_a-\ft12X\sum_{i=1}^NR_i^a
\label{zetanew}\ee
where the equality requires the use of the relationships \eqn{charges} 
and \eqn{mirrinv}. Finally, we are left with,

\paragraph{}

{\bf Theory A:} ${\cal N}=2$ $U(1)^r$ gauge theory with $N$ chiral 
multiplets of charge $R^a_i$, $i=1,\cdots,N$ and $a=1,\cdots,r$ with 
CS parameters $\kappa^{ab}=\ft12 R^a_iR^b_i$, FI parameters $\zeta^{\prime a}$, 
and mass parameters $m_i^\prime$.

\paragraph{}

The prescription to compute the mirror theory is clear. Firstly one should 
perform the mirror rescaling of the parameters. For the masses the mirror 
of \eqn{zetanew}, is 
simply $\hat{m}_i^\prime=\hat{m}_i-\ft12 X$. However, the rescaling of the 
FI parameters is a quantum 
effect, arising from integrating out the chiral multiplets. The correct 
rescaling is induced by a  shift of the vector multiplet scalars, 
$\hat{\phi}^\prime_p=\hat{\phi}_p-\ft12 X$. After weakly 
gauging $U(1)_{N-R}$, the masses 
of the hypermultiplets in Theory B are given in terms of these rescaled 
parameters as
\be
\sum_{i=1}^N(S_i^p\hat{\phi}^\prime_p+\hat{m}_i)^2|{q}_i|^2 
+\sum_{i=1}^N(-S_i^p\hat{\phi}^\prime_p-\hat{m}_i-2X)^2|\tilde{q}_i|^2
\label{pot8}\ee
Integrating out the $\tilde{q}_i$ as $X\rightarrow \infty$ induces the CS parameters 
$\hat{\kappa}^{pq}=-\ft12 S_i^pS_i^q$, together with the finite renormalisation 
of the FI parameters, 
$\hat{\zeta}^p\rightarrow\hat{\zeta}^{\prime p}= \zeta^p +\ft12 
X\sum_{i=1}^N S_i^p$, which is indeed the mirror deformation of \eqn{massnew}. 
Thus the final result is,

\paragraph{}

{\bf Theory B:} ${\cal N}=2$ $U(1)^{N-r}$ gauge theory with $N$ chiral 
multiplets of charge $S^p_i$, $i=1,\cdots,N$ and $p=1,\cdots,N-r$, with 
CS parameters $\hat{\kappa}^{pq}=-\ft12 S_i^pS_i^q$, FI parameters 
$\hat{\zeta}^{\prime p}$ and mass parameters $\hat{m}_i^\prime$.

\paragraph{}

This theory is mirror to Theory A when the masses and FI parameters 
are related through the relevant mirror invariant quantities \eqn{mirrinv}. 
It was shown in \cite{dt} that the Coulomb branch of Theory A 
is equivalent as a toric variety to the Higgs branch of Theory B and 
vice versa. 
The latter is an $r$-dimensional complex space given by the 
usual symplectic quotient construction. The Coulomb branch 
is more interesting: the requirement that the effective CS parameter 
vanishes after integrating out the remaining chiral multiplets restricts 
$\phi$ variables to lie within a region $\Delta\subset {\bf R}^r$. The dual 
photons provide a torus ${\bf T}^r$ which is fibered over $\Delta$ 
such that certain cycles shrink at the boundaries. The resulting space 
is equivalent to the Higgs branch. 

\subsection*{An Example: ${\bf CP}^2$}

Let us illustrate the construction of the Coulomb branch as a toric 
variety  with 
a particularly simple example. For further details, including a 
description of the most general case, see  \cite{dt}. 
It is well known how to construct 
${\bf CP}^2$ as the Higgs branch of 
Theory A: we must take a single $U(1)$ gauge factor with 3 chiral 
multiplets, each of charge $+1$. Setting the masses to 
zero ensures that the vacuum moduli space is given by the vanishing of 
the D-term, 
\be
\sum_{i=1}^3|q_i|^2=\zeta
\nn\ee
which, after modding out by gauge transformations $q_i\rightarrow 
e^{i\alpha}q_i$, leads to the promised projective space. 

Before turning to Theory B, let us first describe ${\bf CP}^2$ in 
more detail, following the discussion of \cite{vl}. 
At a generic point, ${\bf CP}^2$ admits an action of ${\bf T}^2$, 
parametrised by the periodic coordinates $\theta_1$ and $\theta_2$.  
We may choose a basis of this to be
\be
(q_1,q_2,q_3)\rightarrow(e^{i\theta_1}q_1,e^{i(\theta_2-\theta_1)}q_2, 
e^{-i\theta_2}q_3)
\nn\ee
It is interesting to look at the fixed points of this action. There 
are three combinations of cycles which have fixed points: the 
cycle $\theta_1+2\theta_2$ degenerates on the ${\bf CP}^1$ submanifold 
$q_1=0$; the cycle $\theta_2+2\theta_1$ degenerates on the 
${\bf CP}^1$ submanifold $q_3=0$; the cycle $\theta_1-\theta_2$ 
degenerates on the ${\bf CP}^1$ submanifold $q_2=0$. There are also 
three fixed points where two out of three of these cycles vanish. 
These observations allow us to represent ${\bf CP}^2$ as shown in 
Figure 2.

\begin{figure}
\begin{center}
\epsfxsize=2.5in\leavevmode\epsfbox{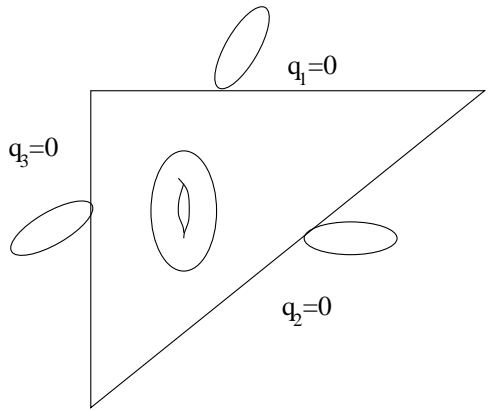}
\end{center}
\caption{\em The toric realisation of ${\bf CP}^2$. A torus ${\bf T}^2$ is 
fibered over the triangle such that a single cycle degenerates at each 
edge. Each of these edges is itself a copy of ${\bf CP}^1$ of the form 
shown in Figure 1.}
\label{fig2}
\end{figure}

Let us now turn to the Coulomb branch of Theory B which, in this 
case is $U(1)^2$ with 3 chiral multiplets of charge $S_i^p$ given 
by $(1,0)$, $(-1,1)$ and $(0,-1)$. Each has mass $m=\ft23\zeta$. 
As described above, the bare CS coupling is given by 
$\kappa^{pq}=-\ft12 S_i^pS_i^q$, but on the Coulomb branch, it also 
receives contributions from 
the remaining chiral multiplets, so that the effective CS parameter 
is given by,
\be
\kappa^{pq}=\ft12 \sum_{i=1}^3S_i^pS_i^q\left(-1+{\rm sign}(M_i)\right)
\nn\ee
where $M_i$ is the effective mass of the $i{}^{\rm th}$ chiral 
multiplet, given by 
\be
M_1&=&\hat{\phi}_1+m \nn\\
M_2&=&-\hat{\phi}_1+\hat{\phi}_2 +m \nn\\
M_3&=&-\hat{\phi}_2+m
\nn\ee
The Coulomb branch exists for $\kappa^{pq}=0$, which requires 
${\rm sign}(M_i)=+1$ for $i=1,2,3$, and therefore restricts the 
Coulomb branch to the region shown in Figure 3.

\begin{figure}
\begin{center}
\epsfxsize=3.0in\leavevmode\epsfbox{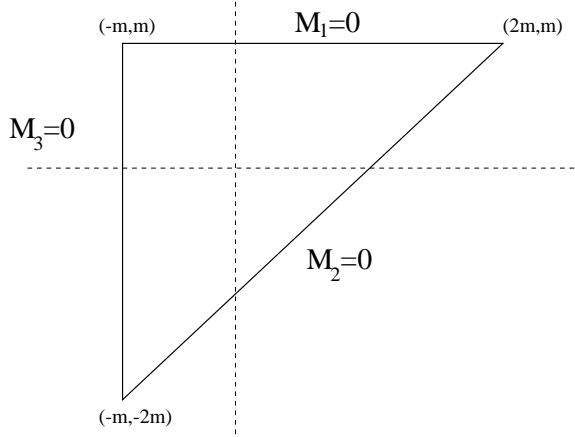}
\end{center}
\caption{\em The Coulomb branch of Theory B. Outside the triangle, 
CS terms develop, lifting the Coulomb branch. The two dual photons 
provide a torus ${\bf T}^2$ which is fibered over the triangle to 
realise ${\bf CP}^2$ as shown in Figure 2.}
\label{fig3}
\end{figure}

The remaining part of the Coulomb branch arises from the two dual 
photons, $\sigma_1$ and $\sigma_2$ which supply a two torus ${\bf T}^2$ 
at generic points of the Coulomb branch. However, at the boundary 
of the Coulomb branch, the chiral multiplet $M_i$ becomes massless 
and one of the cycles of the ${\bf T}^2$ shrinks. In order to see 
which cycle is shrinking, note that when $M_i=0$, a Higgs branch 
emerges from the boundary if we gauge the $U(1)_J$ current 
that induces a constant shift of the combination $S_i^a\sigma_a$ \cite{dt}. 
This therefore is the shrinking cycle. Relating the phases that we introduced 
on the Higgs branch with the dual photons via
\be
\left(\begin{array}{c} \sigma_1 \\ \sigma_2 \end{array}\right) = 
\frac{1}{\sqrt{3}}\left(\begin{array}{cc} 1 & 2 \\ -1 & 1 \end{array}
\right)\left(\begin{array}{c} \theta_1 \\ \theta_2 \end{array}\right)
\nn\ee
completes the identification of Coulomb and Higgs branches. 

\subsection*{Magnetic Coupling}

The mirror map exchanging mass with FI parameters relates the global 
symmetries of the two theories. In particular, weakly gauging the global flavour 
symmetry of Theory A introduces mass parameters. The mirror deformation, 
namely introducing FI parameters, can be achieved by weakly gauging the  
global $U(1)_J$ symmetries of Theory B. 

In \cite{ks}, Kapustin and Strassler pointed out that one needn't be 
so weak when gauging these symmetries. In other words, the newly 
introduced coupling constants may be kept finite and mirror 
symmetry still holds. The gauge potential of the flavour symmetries 
couples to the hypermultiplets in the standard, electric, fashion, with 
coupling constant $e$. However, the $U(1)_J$ symmetries couple to the 
hypermultiplets through a BF-coupling (dubbed four-fermi coupling in 
\cite{ks}). The resulting coupling constant, $\tilde{e}$, is often 
referred to as the ``magnetic coupling''. Although this terminology 
is somewhat misleading, it is widespread in the literature so we 
retain it here with the caveat that this is not the electric-magnetic 
dual coupling - see \cite{ks}.  

Of course, we know only the relationship between 
the Cartan subalgebras of the global 
symmetries: the full non-abelian symmetry of the Coulomb branch is 
not seen in the classical Lagrangian. This limits our identification of 
this magnetic coupling to abelian gauge theories. Presumably, a better 
understanding 
of enhanced global symmetries in these theories may suggest a way 
to extend this technique to non-abelian gauge theories as well. 

In this subsection, we merely comment that the operations described 
above are compatible 
with the partial breaking of supersymmetry.
For example, we may consider the simple ${\cal N}=4$  model of 
$U(1)$ with a single hypermultiplet and finite coupling constant $e$. 
Weakly gauging the $U(1)_{R-N}$ symmetry of this model as described 
above leads to ${\cal N}=2$ $U(1)$ gauge theory with a single 
chiral multiplet of charge $+1$ and a CS parameter $\kappa = +\ft12$. 
The resulting Coulomb branch exists only for $\phi\leq 0$ and has 
one-loop metric
\be
{\bf d}s^2 = \left(\frac{1}{e^2}+\frac{1}{2|\phi |}\right)
{\rm d}\phi^2 + \left(\frac{1}{e^2}+\frac{1}{2|\phi |}\right)^{-1}
{\rm d}\sigma^2\ \ \ \ \ \ \ \ \ (\phi\leq 0)
\label{kquot}\ee
On the other side, the mirror ${\cal N}=4$ theory is given by 
$U(1)_1\times U(1)_2$ gauge group with a single 
hypermultiplet which is coupled only to the 
first $U(1)$ factor. While the gauge coupling $e_1$ is sent to infinity, 
$e_2$ may be kept finite and is to be identified with the coupling 
$e$ of Theory A under mirror symmetry.   
The two vector multiplets are coupled through a BF 
term\footnote{Strictly speaking $U(1)_2$ is a twisted 
vector multiplet. This distinction no longer exists in ${\cal N}=2$ 
theories. See \cite{ks} for further details.}. 
Weakly gauging $U(1)_{N-R}$ of this theory results 
in the ${\cal N}=2$ theory with $U(1)^2$ gauge group and a single 
chiral multiplet of charge $(+1,0)$. The bare CS coupling is given 
by
\be
\kappa^{ab}=\left(\begin{array}{cc}-\ft12 & +\ft12 \\ +\ft12 & 0 
\end{array}\right) 
\nn\ee
where the off-diagonal terms are the ${\cal N}=2$ relic of the 
${\cal N}=4$ BF-coupling and $\kappa^{11}$ 
is induced from integrating out the chiral multiplet. The scalar potential 
of this theory is given by,
\be
U=e_1^2(|q|^2-\ft12\hat{\phi}_1-\hat{\phi}_2)^2 
+ e_2^2(\hat{\phi}_1)^2 +\hat{\phi}_1^2|q|^2
\nn\ee
which has a Higgs branch given by $\hat{\phi}_1=0$, with $q$ and 
$\hat{\phi}_2$ constrained to satisfy the first D-term, modulo 
$U(1)_1$ gauge transformations. Importantly, the dual photon $\hat{\sigma}_2$ 
transforms as $\hat{\sigma}_2\rightarrow\hat{\sigma}_2+\alpha$ under 
such transformations \cite{ks}. It may be checked explicitly that 
the resulting K\"ahler quotient reproduces the metric \eqn{kquot} above. 

Finally note that, in the spirit of \cite{ks}, we may also consider keeping 
a finite coupling constant for the gauged $U(1)_{R-N}$ symmetry, 
preserving the property of mirror symmetry.

\section{Non-Abelian Chern-Simons Theories}

The idea of gauging the R-symmetry to flow from ${\cal N}=4$ 
to ${\cal N}=2$ theories applies equally well to non-abelian 
mirror pairs, of which many are known. Here we will discuss the 
dynamics of the class of theories that arise in such a construction.

Let us recall a few relevant facts concerning the non-abelian CS coupling, 
\be
{\cal L}_{CS}=\frac{\kappa}{4\pi}\epsilon^{\mu\nu\rho}
{\rm Tr}\left(A_\mu\partial_\nu A_\rho+\ft23 A_\mu 
[A_\nu,A_\rho]\right)
\label{csna}\ee
Unlike in the abelian case, the action is not gauge invariant. 
The requirement of invariance of the partition function under 
gauge transformations that that are homotopically non-trivial in 
$\pi_3(G)$ requires the CS parameter to be quantised: 
$\kappa\in Z$. For non-abelian groups, gauge invariance 
does not allow for the possibility of cross CS terms of the form 
\eqn{crosskappa}.

In non-supersymmetric pure Yang-Mills Chern-Simons theories, 
the CS parameter $\kappa$ undergoes a finite, integer 
renormalisation \cite{sumati}. 
This is not the case in the ${\cal N}=2$ theories considered 
here \cite{kimyeong}. However, as before, $\kappa$ is 
renormalised by integrating out massive chiral 
multiplets. Specifically, consider 
$N$ chiral multiplets of real mass $M_i$, transforming in the 
representation $R_i$ of some unbroken semi-simple gauge group $G$. 
Integrating out this matter results in the effective 
CS parameter
\be
\kappa \rightarrow \kappa+\ft12\sum_{i=1}^Nd_2(R_i)\,{\rm sign}(M_i)
\label{nacs}\ee
where $d_2(R_i)$ is the quadratic casimir of $R_i$ normalised such that 
$d_2({\bf N})=1$ for the fundamental representation of $SU(N)$. 
Notice in particular, the factor of $\ft12$ in 
\eqn{nacs} implies that an ${\cal N}=2$ 
$SU(N)$ theory with a single chiral multiplet in the fundamental 
representation is consistent only if one adds a bare half-integer 
CS parameter. Such a term breaks parity at the classical level and 
is referred to as a parity anomaly \cite{redlich,agw}. As in the 
abelian case, it is possible for parity to be restored in the 
effective theory after integrating out the chiral multiplets. 

To illustrate mirror symmetry, let us consider the simplest non-abelian 
mirror pairs discovered by Intriligator and Seiberg \cite{intseib}

\paragraph{}

{\bf Theory A:} ${\cal N}=4$ $SU(2)$ gauge theory with 
$N\geq 4$ hypermultiplets in the ${\bf 2}$ representation.

\paragraph{}

{\bf Theory B:} ${\cal N}=4$ $D_N$ quiver theory with gauge group 
\be
G_{D_N}=\prod_{a=1}^{N+1}U(n_a)/U(1)
\nn\ee
where $n_a=1$ for 
$a=1,2,3,4$ and $n_a=2$ for $a=5,\cdots ,N+1$ are the Dynkin indices 
of the $D_N$ diagram. Bi-fundamental hypermultiplets transform in the 
$c_{ab}({\bf n}_a,\bar{\bf n}_b)$ where $c_{ab}=1$ if a link 
joins the $a^{\rm th}$ and $b^{\rm th}$ node, and is zero otherwise 
- see \cite{intseib} for further details.

\paragraph{}

Gauging the $U(1)_{R-N}$ symmetry to introduce an axial mass  
proceeds in the same manner as in the previous section. The 
only difference is that the adjoint chiral superfield $\Psi$ now also 
contributes to the CS parameter. For $SU(2)$ gauge group, the casimirs are 
related by $d_2({\bf 3})=4d_2({\bf 2})$. After a suitable shift of the 
mass parameters, Theory A flows to 

\paragraph{}

{\bf Theory A:} ${\cal N}=2$ $SU(2)$ gauge theory with $N$ chiral 
multiplets in the ${\bf 2}$ representation with non-abelian 
CS parameter $\kappa=\ft12 N-2$.

\para

For the quiver theory, each $SU(2)\subset G_{D_N}$ has four fundamental 
hypermultiplets. After a suitable shift of the $U(1)$ scalars, the 
fundamental and adjoint chiral multiplet cancel in their contribution 
to the non-abelian CS parameter. The abelian CS parameters do receive contributions 
however. We find,

\para

{\bf Theory B:} ${\cal N}=2$ $D_N$ quiver theory, with gauge group 
$G_{D_N}$ with bi-fundamental chiral multiplets determined by the 
connections of the $D_N$ Dynkin diagram. 
The non-abelian CS parameter is zero, while the abelian CS parameters are given 
by $\kappa_{aa}=-1$ for $a=1,2,3,4$ and  $\kappa_{aa}=-4$ for 
$a=5,\cdots,N+1$ and $\kappa_{ab}=+\ft12 c_{ab}n_an_b$ if $a\neq b$.

\para

Rather than enter into the details of elucidating agreement between the 
vacuum structure of these two theories, we will instead concern ourselves 
with a discussion of the class of theories that arise, namely non-abelian 
gauge theories with chiral multiplets. 

\subsection*{SU(2) Theories}

We start by discussing the simplest non-abelian theory:  
${\cal N}=2$ $SU(2)$ gauge theory with 
zero bare CS parameter. The vector multiplet includes a real scalar 
$\phi$ in the adjoint representation. We further include two 
chiral multiplets in the ${\bf 2}$ representation with real masses $-m$ and $+M$ 
where we choose $m,M>0$.

Integrating out the massive chirals will result in CS couplings. Let us first 
position ourselves far out on the Coulomb branch, breaking 
$SU(2)\rightarrow U(1)$ with $\phi=v\tau^3$, where $\tau^3$ is the third Pauli 
matrix. The effective description is in terms of the surviving $U(1)$ gauge 
group. 
Thus the non-abelian CS couplings \eqn{csna} are not relevant in determining the 
vacuum structure and we must look once more to the now-familiar abelian CS 
coupling. 

Each chiral multiplet, $q_i$ (with $i=1,2$ a flavour index) decomposes 
into two chirals $q_i^a$ under the unbroken $U(1)$ (with $a=1,2$ a colour 
index). The chirals 
$q_i^1$ have charge $+1$ while $q_i^2$ have charge $-1$. 
Their masses are given by,
\be
(v-m)^2|q^1_1|^2 +(-v-m)^2|q^2_1|^2+(v+M)^2|q_2^1|^2+(-v+M)^2|q^2_2|^2
\nn\ee
Notice that if $M=m$ then the chiral multiplets come in pairs with 
opposite real masses. In such a case, no CS parameter will be 
generated. We will therefore restrict attention to a somewhat different limit: 
$M\rightarrow\infty$ with $v$ and $m$ kept finite. 
Integrating out $q_2^a$ then generates an abelian CS coupling with 
$\kappa=\ft12\times 2\times{\rm sign}(M)=+1$. We must also integrate out 
the W-boson multiplets. However the these come with opposite charges 
and opposite real masses induced by the Higgs mechanism and so do not 
contribute to $\kappa$. 
Following the discussion in the previous section, we see that the 
Coulomb branch exists if, after integrating out the remaining 
chiral multiplets $q_1^a$, the effective CS parameter vanishes,
\be
\kappa=+1+\ft12\,{\rm sign}(v-m)+\ft12\,{\rm sign}(-v-m)=0
\nn\ee
which requires $|v|<m$. This is the same restriction we found 
in the $U(1)$ theory with two chirals. This, of course, is not 
surprising as, after the Higgs mechanism, the matter content coincided 
with the abelian model. However, the dynamics of this theory remember 
their non-abelian origin, and therefore differ from the $U(1)$ theory, in 
three distinct ways. Firstly, the residual Weyl symmetry allows 
us to fix $0<v<m$. Secondly, the description of the Coulomb branch 
in terms of a periodic scalar fibered over the interval breaks down 
at $v=0$ where non-abelian gauge symmetry is classically restored 
and we cannot perform the manoeuvres necessary to dualise the 
gauge field. The resulting perturbative Coulomb branch is 
shown in Figure 4.

\begin{figure}
\begin{center}
\epsfxsize=3.0in\leavevmode\epsfbox{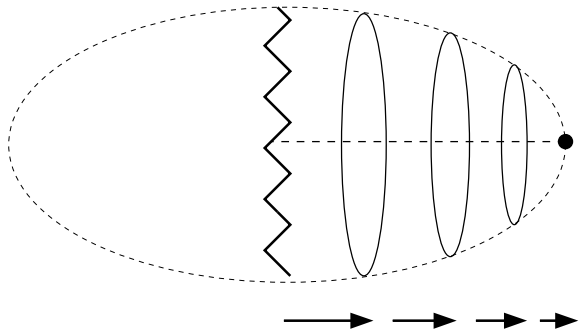}
\end{center}
\caption{\em The Coulomb branch of the $SU(2)$ theory in the 
limit $M\rightarrow\infty$. Non-abelian 
gauge symmetry is restored at the origin, denoted by the 
vertical jagged line. Instantons, represented by the arrows, induce 
a potential driving the vacuum to the point marked by the 
solid dot. This theory has a unique supersymmetric vacuum.}
\label{fig4}
\end{figure}

In fact, in this model the troublesome point at $v=0$ is actually removed from 
consideration by the third difference: instantons. In three dimensional 
gauge theories these are magnetic monopole configurations. 
Explicit instanton calculations have been performed in three 
dimensional $SU(2)$ theories with no supersymmetry \cite{poly} as well 
as with ${\cal N}=2$ \cite{afw,dtv}, ${\cal N}=4$ \cite{dkmtv,dtv} 
and ${\cal N}=8$ \cite{dkm} supersymmetry. Discussions of their 
effects in ${\cal N}=2$ theories with hypermultiplets can be found in 
\cite{berk3,ahiss}. More recently (last week) there has also been a discussion 
of the effects of instanton-anti-instanton pairs in ${\cal N}=4$ theories 
\cite{ami}. 

Let us recall a few simple facts. In ${\cal N}=2$ theories, instantons 
may generate a superpotential. To determine whether or not such a 
potential does indeed arise in a given case, one must calculate the 
instanton contribution to the two-fermi correlation function 
$\langle{\rm Tr}\lambda^2\rangle$, where $\lambda$ are the vector multiplet 
fermions. This is non-zero only if the instanton has precisely 
two unlifted fermionic zero modes. Thus we need determine the number of 
fermionic collective coordinates of a given instanton configuration.  
These may arise from either vector multiplet or chiral multiplet fermions. 
The surviving two must be of vector multiplet origin. 

The counting of the zero modes follows from the Callias index theorem 
\cite{callias}. A charge $k$ instanton 
has $2k$ fermionic zero modes donated by the vector multiplet. 
In contrast, a chiral multiplet of real mass 
$m$ provides $k$ fermionic zero modes only for $|m|<v$. Otherwise, there are 
no zero modes.

In the case at hand, the perturbative Coulomb branch is restricted 
to $0<v<|m|$. This is precisely the region in which the chiral multiplet 
fermions provide no zero modes.  
We therefore find only the $2k$ vector multiplet zero 
modes and expect a superpotential to be generated by 
a charge one instanton. Indeed, one can explicitly calculate the 
contribution to the superpotential from a single instanton,
\be
{\cal W} = ce^2\exp (-Y) \sim (c/e^2)v^3\sqrt{\frac{|v-m|}{|-v-m|}}\, 
{\rm exp}(-v/e^2+i\sigma)
\label{spot}\ee
where we have introduced $Y$, a holomorphic coordinate on the 
Coulomb branch, as well as the constant $c$. This constant has 
been calculated explicitly in \cite{dtv} and is non-zero. 

The second relation in \eqn{spot} is interesting in its own right 
and we digress somewhat here to explain its significance. It may 
be determined in two different ways. In the first of these approaches, 
one calculates the one-loop correction 
to the complex structure on the Coulomb branch. This determines the 
corrections to the classical relationship $Y=v/e^2 +i\sigma$. 
Holomorphicity of the superpotential then ensures that it takes the above 
form when expressed in terms of the microscopic variables \cite{berk3}. 
The $\sim$ sign in \eqn{spot} means ``up to two-loop perturbative corrections''. 
In the second, more direct approach, the factor in front of the exponent 
arises through a calculation of one-loop determinants around the background of 
the instanton. While in four dimensions, such effects famously cancel in 
supersymmetric theories \cite{thooft}, the same is not true in three 
dimensional theories  with less than sixteen supercharges \cite{dkmtv}. An 
explicit calculation of these terms in ${\cal N}=2$ theories with 
matter was performed in \cite{dtv}. The $\sim$ sign now means 
``up to two-loops around the background of the instanton''. It 
is intriguing to note that in these theories holomorphicity relates 
complicated perturbative effects around the background of the instanton 
to perturbative effects in the vacuum.

In the present situation, the factor in front of the exponent has 
an important role to play: it ensures the existence of an 
isolated supersymmetric vacuum. To see this, first recall that 
supersymmetric vacua exist if, 
\be
\frac{\partial {\cal W}}{\partial Y} =0 
\iff {\cal W}=0
\nn\ee
where the ``if and only if'' refers to the particular superpotential \eqn{spot}. 
We therefore find a supersymmetric vacuum state at $v=m$. 
This is shown in Figure 4. In this vacuum, the chiral multiplet 
becomes massless and is constrained by a quartic superpotential. It is 
possible that this point corresponds to an interacting ${\cal N}=2$ 
superconformal theory.

\subsubsection*{\it Variations on a Theme}

One may consider adding various extra  matter multiplets to 
this $SU(2)$ model. For example, the mass $M$ of the second 
chiral could be kept finite. This then results in disconnected 
components of the Coulomb branch at $v\geq {\rm max}(m,M)$ 
and $v\leq{\min}(m,M)$. In the first of these components, 
no superpotential is generated. In the second, a superpotential 
similar to \eqn{spot} drives the vacuum to $v={\min}(m,M)$ 
where the superpotential vanishes.

Suppose further that one added a single hypermultiplet (i.e. 
two chiral multiplets with opposite real mass) 
in the ${\bf 2}$ of the gauge group. We send $M\rightarrow\infty$ once more and set 
the hypermultiplet mass to be equal to 
$\alpha m$ where $0\leq\alpha\leq 1$. While this 
hypermultiplet does not alter the perturbative Coulomb branch 
(it does not generate CS terms), it does affect the superpotential. 
Indeed, it contributes $2k$ fermionic zero modes to the charge 
$k$ instanton whenever $v>\alpha|m|$. The superpotential is 
therefore only generated in the regime $v<\alpha |m|$, and part 
of the Coulomb branch survives in the quantum theory. Although 
this appears to violate holomorphy of the superpotential, it 
was shown in \cite{berk3, ahiss} that the Coulomb branch splits 
about the point $v=\alpha m$. This phenomenon is related to the 
shrinking of the toric fibres in the discussion of the Coulomb branch 
in the previous section. 

There is also a 
two dimensional Higgs branch 
extending from the point $v=\alpha m$. It was argued in \cite{ahiss} 
that the Higgs and the Coulomb branches merge smoothly at the quantum 
level. The same behaviour occurs here, the difference being that the 
Coulomb branch no longer stretches to infinity.

\subsection*{SU(N) Theories}

We now repeat the analysis of the previous section for 
$SU(N)$ theories with 2 chiral multiplets transforming in 
the ${\bf N}$ representation. Again, we endow these multiplets 
with masses $-m$ and $+M$ where $m,M\geq 0$. Unlike the 
$SU(2)$ model, we will see a somewhat richer behaviour as the 
mass parameters are varied.

Let us first perform the necessary group theory decomposition that 
will be needed in both cases. We proceed by first moving onto the Coulomb branch 
by assigning a vacuum expectation value 
$\phi ={\bf v}\cdot{\bf H}={\rm diag}(v_1,\cdots,v_N)$ to the 
adjoint scalar. Here ${\bf v}$ is a $(N-1)$-vector and ${\bf H}$ denotes the 
generators of the Cartan subalgebra. We will find it at times 
useful to alternate between Cartan-Weyl notation and the more explicit 
notation in which $\sum_{i=1}^N v_i=0$. 

We assume that ${\bf v}$ is such that $SU(N)$ breaks to the maximal torus, 
$U(1)^{N-1}$. Each 
chiral multiplet decomposes into $N$ chiral multiplets transforming 
under the surviving abelian gauge groups, with charges given by 
$\bomega_i\cdot\bbeta_aC^{ab}$ where ${\bomega_i}$, $i=1,\cdots,N$ are the 
fundamental weights of $SU(N)$, $\bbeta_a$, $a=1,\cdots,N-1$, 
are the fundamental roots and $C_{ab}$ is the 
Cartan matrix. The masses 
of these chiral multiplets are 
given by $\bomega_i\cdot{\bf v}-m=v_i-m$ and 
$\bomega_i\cdot{\bf v}+M=v_i+M$.  Moreover, in the case of maximal symmetry 
breaking we may exchange each of the $N-1$ photons for a dual 
scalar $\sigma_a$.

Integrating out the chiral multiplets yields,
\be
\kappa_{ab}=\ft12\sum_{i=1}^N(\bomega_i\cdot\bbeta_a)
(\bomega_i\cdot\bbeta_b)\left({\rm sign}\,(v_i-m)+{\rm sign}(v_i+M)\right) 
\nn\ee
where all $a$ and $b$ indices are to be raised and lowered by the 
Cartan matrix. Inserting explicit expressions for the weights and 
roots, we have
\be
\kappa_{ab}&=&\ \left[{\rm sign}(v_a-m)+{\rm sign}(v_a+M)+{\rm sign}(v_{a+1}-m)
+{\rm sign}(v_{a+1}+M)\right]\delta_{a,b} \nn\\
&&-\left[{\rm sign}(v_a-m)+{\rm sign}(v_a+M)\right]\delta_{a,b+1} \nn\\ 
&&-\left[{\rm sign}(v_{a+1}-m)+{\rm sign}(v_{a+1}+M)\right]\delta_{b,a+1}
\nn\ee
The existence of a Coulomb branch again requires $\kappa_{ab}=0$, a 
requirement which 
is strongest for the off-diagonal terms, $a=b\pm 1$. These 
give the condition $-M\leq v_i\leq m$ for each   
$i=1,\cdots,N$. Upon fixing the Weyl group, the perturbative Coulomb branch is 
therefore restricted to 
\be
m\geq v_1\geq v_2\geq\cdots\geq v_N = -\sum_{i=1}^{N-1} v_i \geq -M
\label{suncol}\ee
The perturbative Coulomb branch for $SU(3)$ gauge group in the limit 
$M\rightarrow\infty$ is shown in Figure 5. 

\begin{figure}
\begin{center}
\epsfxsize=3.0in\leavevmode\epsfbox{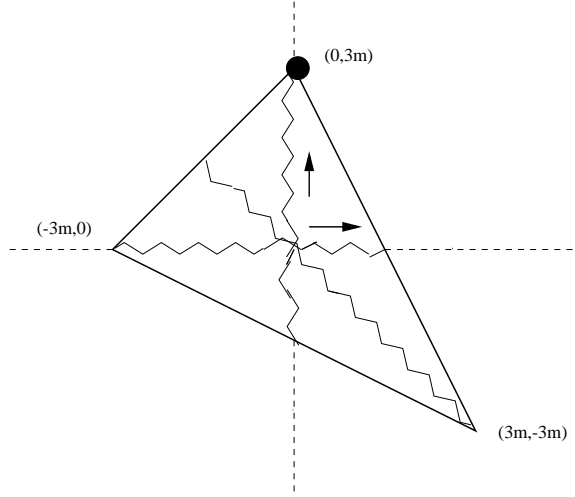}
\end{center}
\caption{\em The Coulomb branch of the $SU(3)$ theory with 
$M\rightarrow\infty$. The horizontal and vertical lines are 
$v_1-v_2$ and $v_2-v_3$ respectively. The jagged lines lie at 
the edge of the Weyl chambers and denote non-abelian symmetry 
enhancement. Instantons carry the unique supersymmetric 
vacuum to the spot marked with a dot.}
\label{fig5}
\end{figure}   

We must now analyse instanton effects in this theory. 
We again recall some salient facts about monopoles in higher 
rank gauge groups. The topological charge of such objects is given by 
a linear combination of co-roots:  
${\bf g}=\sum_{a=1}^{N-1}k_a\bbeta_a^{\star}$ with $k_a\in Z$. Such a 
monopole  has $2\sum_{a=1}^{N-1}k_a$ fermionic zero modes \cite{weinberg} 
from vector multiplets.  As we are interested 
in superpotential-generating-instantons, we restrict ourselves to the 
$N-1$ ``fundamental'' sectors given by ${\bf g}=\bbeta_a^\star$, each 
of which have the requisite two fermi zero modes. The story with 
chiral multiplets is a little more complicated. Having fixed the 
Weyl symmetry as above, a chiral multiplet of mass real $m$ such that 
$v_i\geq m \geq v_{i+1}$ donates a single fermi zero mode to only 
the $i^{\rm th}$ fundamental monopole. In particular if, as is 
the case in \eqn{suncol}, $m\geq v_i$ or $m\leq v_i$ for all $i$, 
then it contributes no zero modes. Thus we see that all $(N-1)$ fundamental 
monopoles can contribute to the superpotential.

Explicit calculations 
of instanton contributions in $SU(N)$ ${\cal N}=4$ and ${\cal N}=8$ 
theories, including one-loop factors, were performed in \cite{ft} and 
can be easily generalised to the case at hand. Discussions of the 
superpotentials generated in ${\cal N}=2$ $SU(N)$ theories are given in 
\cite{berk3, ahiss}. The superpotential is,
\be
{\cal W}&=&ce^2\sum_{a=1}^{N-1}\exp (-Y_a) \label{bigspot}\\
&\sim& (c/e^2)\sum_{a=1}^{N-1}(\bbeta_a\cdot {\bf v})^3
\sqrt{\frac{|v_a-m|}{|v_{a+1}-m|}}\sqrt{\frac{|v_a+M|}{v_{a+1}-M|}}
\exp(\bbeta_a\cdot{\bf v}/e^2+i\sigma_a)
\nn\ee
where $c$ is a non-zero constant and $\sim$ means once again ``up to two-loop effects''. 
The one-loop determinants which make up the prefactor in front of 
the exponent may be extracted from \cite{berk3} or \cite{ft,dtv}.

The form of the superpotential once again ensures that the scalar 
potential vanishes if ${\cal W}=0$. We therefore 
require $v_a=+m$ or $v_a=-M$ for $a=1,\cdots,N-1$. Let us choose,
\be
v_a&=&+m\ \ \ \ \ \ \ \, a=1,\cdots,r \nn\\
v_a&=&-M\ \ \ \ \ \ \ a=r+1,\cdots,N-1 \label{vacuum}\\
v_N&=&-\sum_{a=1}^{N-1}v_a=-rm+(N-r-1)M
\nn\ee
Notice that whenever the denominator of \eqn{bigspot} blows up, 
the term $({\bf v}\cdot\bbeta_a)^3$ vanishes faster and the point \eqn{vacuum} 
is indeed a zero of ${\cal W}$.

We are now in a position to see the fate of this theory. First, 
let us note that, for certain values of the mass parameters, 
the critical points of the superpotential \eqn{vacuum} may not 
lie within the perturbative Coulomb branch \eqn{suncol} and do    
not therefore necessarily correspond to vacua of the theory. 
Let us examine the conditions under which this occurs. The 
troublesome inequality is that on the right-hand-side of \eqn{suncol} 
when combined with the final equation of \eqn{vacuum}. Generically, 
this latter equation requires $r=N-1$, and it is possible to satisfy the  
inequality only if
\be
M\geq (N-1)m
\nn\ee
in which case the unique vacuum state lies at $v_a=m$ for $a=1,\cdots,N-1$ 
and $v_N=(1-N)m$. For $SU(3)$ gauge group, this vacuum is shown in 
Figure 5. Rather surprisingly, the non-abelian gauge 
symmetry is partially 
restored in this vacuum. Specifically, semi-classical analysis suggests 
that $SU(N)\rightarrow SU(N-1)\times U(1)$. However, the dynamics at this 
point is strongly coupled and the resulting physics is unclear: is there 
a mass gap or a non-trivial conformal theory? Clearly it would be interesting 
to understand these features better. 

If $M<(N-1)m$, then the perturbative Coulomb branch contains no 
critical points of the superpotential. In this case, either 
supersymmetry is broken or the K\"ahler potential is singular, 
yielding a supersymmetric vacuum. While the presence of the 
non-abelian enhancement point may suggest the latter resolution, 
it is noteworthy that dynamical supersymmetry breaking has been 
seen previously in ${\cal N}=1,2$ and $3$ 
Chern-Simons theories using brane techniques \cite{csbreking, andreas}. 
For ${\cal N}=1$ theories, the Witten index has also been explicitly 
computed \cite{wittd}.

For completeness we note the existence of extra 
potential supersymmetric vacua \eqn{vacuum} whenever $r=NM/(m+M)\in Z$. 
Such a vacuum has $SU(r)\times SU(N-r-1)\times U(1)$ gauge symmetry. 
Again, the physics here is poorly understood.

Finally, we briefly mention that one can repeat this analysis for the 
related $U(N)$ theory with two chiral multiplets. This has the 
advantage of being more amenable to brane constructions.  
Indeed, it is simple to construct the $U(N)$ version of the 
above theory on the world-volume of a D3-brane suspended 
between 5-brane webs \cite{andreas,dt}. From the field theory 
side, one finds the same restrictions on the perturbative Coulomb 
branch and the same superpotential. However, unlike the $SU(N)$ 
theory, $v_N$ is not constrained to be equal to the sum of the other vacuum 
expectation values and there is no conflict between the critical 
points of the superpotential and the perturbative Coulomb branch. 
One therefore finds a branch of supersymmetric 
vacua parametrised by $v_N$,  all of which exhibit non-abelian symmetry 
enhancement. 
It is simple to show in the brane picture, using 
the techniques of \cite{ah}, that a generic configuration of 
$D3$-branes will indeed tend towards such a  
vacuum\footnote{I would like to thank Ami Hanany and Andreas Karch 
for several educational discussions on this subject.}.

\section{Summary}

We have shown how to flow from ${\cal N}=4$ theories to 
${\cal N}=2$ theories preserving the mirror map by gauging 
a suitable combination of the R-symmetries. The resulting 
theories have Chern-Simons couplings and chiral multiplets. 
Moreover, we have analysed the dynamics of such theories 
which display interesting phenomena in their own right.  
In particular, for a gauge group of rank $r$, we have:

\para

$\bullet$ The perturbative Coulomb branch is restricted by the 
CS terms to a region $\Delta\subset {\bf R}^r$, which may be 
compact or non-compact depending on the matter content.

\para

$\bullet$ For abelian gauge theories, the dual photons provide a 
torus ${\bf T}^r$ which is fibered over $\Delta$ such that 
certain cycles shrink on the boundaries, resulting in a 
description of the Coulomb branch as a toric variety.

\para

$\bullet$ For non-abelian gauge groups, a dynamically generated 
superpotential forces the vacuum to the boundary of $\Delta$ 
where the gauge symmetry is enhanced to a group of rank $r-1$.

\para

$\bullet$ For non-abelian gauge groups of rank $r\geq 2$, certain 
ranges of the parameters ensure that the critical points of the 
superpotential lie outside the perturbative Coulomb branch 
$\Delta$.

\subsection*{Acknowledgements}

I am grateful to Nick Dorey, Bo Feng, Ami Hanany, 
Kentaro Hori, Ken Intriligator, Boris Pioline, Matt Strassler 
and especially Andreas Karch for 
several useful discussions. Thankyou. Moreover, I am indebted to Ami 
and to all at MIT for their hospitality over the past few months. 
This work is supported in part by funds provided by the U.S. Department 
of Energy (D.O.E) under cooperative research agreement \# DF-FCO2-94ER40818, 
as well as by a PPARC SPG-project grant PPA/G/S/1998/00613. 
I am supported by an EPSRC fellowship.

\end{document}